\documentstyle[12pt]{article}
\newlength{\extraspace}
\newlength{\extraspaces}
\newcommand{\ba}{\begin{eqnarray}
\addtolength{\abovedisplayskip}{\extraspaces}
\addtolength{\belowdisplayskip}{\extraspaces}
\addtolength{\abovedisplayshortskip}{\extraspace}
\addtolength{\belowdisplayshortskip}{\extraspace}}
\newcommand{\ea}{\end{eqnarray}}

\newcommand{\nonu}{\nonumber \\[.5mm]}
\newcommand{\A}{&\!\!\!}

\begin{document}

\thispagestyle{empty}

\hfill \parbox{3.5cm}{hep-th/0306080 \\ SIT-LP-03/05}
\vspace*{1cm}
\vspace*{1cm}
\begin{center}
{\bf Linearizing superon-graviton model(SGM)} \\[20mm]
{Kazunari SHIMA and Motomu TSUDA} \\[2mm]
{\em Laboratory of Physics, Saitama Institute of Technology}
\footnote{e-mail:shima@sit.ac.jp, tsuda@sit.ac.jp}\\
{\em Okabe-machi, Saitama 369-0293, Japan}\\[2mm]
{Manabu SAWAGUCHI} \\[2mm]
{\em High-Tech Research Center, Saitama Institute of Technology}
\footnote{e-mail:sawa@sit.ac.jp}\\
{\em Okabe-machi, Saitama 369-0293, Japan}\\[2mm]
{May 2003}\\[15mm]

%\maketitle

\begin{abstract}
We attempt the linearization of N=1 SGM action describing the nonlinear 
supersymmetric(NLSUSY) gravitational 
interaction of superon(Nambu-Goldstone(N-G) fermion). 
We find that 80+80 field contents may give the off-shell supermultiplet 
of the supergravity(SUGRA)-like 
linearized theory and they are realized explicitly up to $O(\psi^2)$ 
as the composites, 
though they have modified SUSY transformations which closes on super-Poincar\'e(SP) algebra. 
Particular attentions are paid to the local Lorentz invariance in the minimal interaction. 

\end{abstract}
\end{center}

\newpage

In the previous paper\cite{ks1} we have proposed superon-graviton model(SGM) 
of nature. 
SGM is obtained by extending the geometrical arguments of Einstein general relativity theory(EGRT) 
on Riemann spacetime to new (SGM) spacetime where the coset space coordinates $\psi^{i}(i=1,2,\cdots,10)$ of 
${{N=10 \ superGL(4,R)} \over GL(4,R)}$ turning to the N-G fermion degrees of freedom(d.o.f.) 
besides the ordinary Minkowski coordinate $x^{a}$ are attached at every Riemann spacetime point. 
The SGM action of a new Einstein-Hilbert(E-H)-type\cite{ks1}
describes the NLSUSY\cite{va} invariant gravitational interaction of N-G fermion 
$superon$ in Riemann spacetime.   \par
In the last article\cite{sts1}  we have discussed the linearization of the new E-H type action(N=1 SGM action) 
to obtain the equivalent linear(L) SUSY\cite{wzgl} theory in the low energy. 
We have pointed out that 80(bosons)+80(fermions) d.o.f. may give the off-shell 
supermultiplet of the linearized theory.     \par
In this letter we would like to discuss further the the linearization of N=1 SGM. 
Considering a phenomenological potential of SGM, though qualitative and group theoretical, 
discussed in \cite{ks2} based upon the composite picture of LSUSY representation 
and the recent interest in NLSUSY in superstring(membrane) world, 
the linearization of NLSUSY in curved spacetime may be of some general interest. 
Furthermore considering  that the number of the (fundamental) particles 
in SUSY GUTs exceeds 160 and that they are embedded in ${1 \times 2 \times 3}$ gauge structure, 
we are tempted to suppose a certain kind of a internal structure of these particles and/or 
the fundamental nature of spacetime itself which are encoded in the geometry of spacetime.        \par
The linearization of SGM is physically interesting in general, 
even if it were  a existing SUGRA-like theory, 
for the consequent broken LSUSY theory 
is shown to be equivalent 
and gives a new insight into the fundamental structure of 
nature.

For the  self-contained arguments we review SGM action briefly.
SGM action is given by\cite{ks1}; 
\begin{equation}
L_{SGM}=-{c^{3} \over 16{\pi}G}\vert w \vert(\Omega + \Lambda),
\label{SGM}
\end{equation}
\begin{equation}
\vert w \vert = {\rm det}{w^{a}}_{\mu} 
= {\rm det}({e^{a}}_{\mu}+ {t^{a}}_{\mu}),  \quad
{t^{a}}_{\mu}={\kappa^{4}  \over 2i}(\bar{\psi}\gamma^{a}
\partial_{\mu}{\psi}
- \partial_{\mu}{\bar{\psi}}\gamma^{a}{\psi}),
\label{w}
\end{equation} 
where $e^{a}{_\mu}$ is the  vierbein of EGRT, $\psi$ is N-G fermion(superon), 
${\kappa^{4} = ({c^{3}\Lambda \over 16{\pi}G}})^{-1} $ is the fundamental volume of 
four dimensional spacetime of V-A model\cite{va},  
and $\Lambda$ is the  ${small}$ cosmological constant related to the strength of 
the superon-vacuum coupling constant. SGM posesses two mass scales, $G^{-1}$(Planck scale) 
 and ${\Lambda \over G}$($O(1)$). 
$\Omega$ is a new scalar curvature analogous to the Ricci scalar curvature $R$ of EGRT, 
whose explicit expression is obtained  by just replacing ${e^{a}}_{\mu}(x)$  
by ${w^{a}}_{\mu}(x)$ in Ricci scalar $R$. 
These results can be understood intuitively by observing that 
${w^{a}}_{\mu}(x) ={e^{a}}_{\mu}(x)+ {t^{a}}_{\mu}(x)$  defined by 
$\omega^{a}={w^{a}}_{\mu}dx^{\mu}$, where $\omega^{a}$ is the NLSUSY invariant differential forms of 
V-A\cite{va}, is invertible and  $s^{\mu \nu}(x) \equiv {w_{a}}^{\mu}(x) w^{{a}{\nu}}(x)$ 
are a unified vierbein and a unified metric tensor in SGM spacetime\cite{ks1}\cite{st1}. 
The SGM action  (\ref{SGM}) is invariant at least under the following symmetry\cite{st2};
ordinary GL(4R),  
the following new NLSUSY transformation; 
\begin{equation}
\delta^{NL} \psi(x) ={1 \over \kappa^{2}} \zeta + 
i \kappa^{2} (\bar{\zeta}{\gamma}^{\rho}\psi(x)) \partial_{\rho}\psi(x),
\quad
\delta^{NL} {e^{a}}_{\mu}(x) = i \kappa^{2} (\bar{\zeta}{\gamma}^{\rho}\psi(x))\partial_{[\rho} {e^{a}}_{\mu]}(x),
\label{newsusy}
\end{equation} 
where $\zeta$ is a constant spinor and  $\partial_{[\rho} {e^{a}}_{\mu]}(x) = 
\partial_{\rho}{e^{a}}_{\mu}-\partial_{\mu}{e^{a}}_{\rho}$, \\
the following GL(4R) transformations due to (\ref{newsusy});  
\begin{equation}
\delta_{\zeta} {w^{a}}_{\mu} = \xi^{\nu} \partial_{\nu}{w^{a}}_{\mu} + \partial_{\mu} \xi^{\nu} {w^{a}}_{\nu}, 
\quad
\delta_{\zeta} s_{\mu\nu} = \xi^{\kappa} \partial_{\kappa}s_{\mu\nu} +  
\partial_{\mu} \xi^{\kappa} s_{\kappa\nu} 
+ \partial_{\nu} \xi^{\kappa} s_{\mu\kappa}, 
\label{newgl4r}
\end{equation} 
where  $\xi^{\rho}=i \kappa^{2} (\bar{\zeta}{\gamma}^{\rho}\psi(x))$, 
and the following local Lorentz transformation on $w{^a}_{\mu}$; 
\begin{equation}
%\eqalign{
\delta_L w{^a}_{\mu}
= \epsilon{^a}_b w{^b}_{\mu}
%}
\label{Lrw}
\end{equation}
with the local  parameter
$\epsilon_{ab} = (1/2) \epsilon_{[ab]}(x)$    
or equivalently on  $\psi$ and $e{^a}_{\mu}$
\begin{equation}
\delta_L \psi(x) = - {i \over 2} \epsilon_{ab}
      \sigma^{ab} \psi,     \quad
\delta_L {e^{a}}_{\mu}(x) = \epsilon{^a}_b e{^b}_{\mu}
      + {\kappa^{4} \over 4} \varepsilon^{abcd}
      \bar{\psi} \gamma_5 \gamma_d \psi
      (\partial_{\mu} \epsilon_{bc}).
\label{newlorentz}
\end{equation}
The local Lorentz transformation forms a closed algebra, for example, on $e{^a}_{\mu}(x)$ 
\begin{equation}
[\delta_{L_{1}}, \delta_{L_{2}}] e{^a}_{\mu}
= \beta{^a}_b e{^b}_{\mu}
+ {\kappa^{4} \over 4} \varepsilon^{abcd} \bar{\psi}
\gamma_5 \gamma_d \psi
(\partial_{\mu} \beta_{bc}),
\label{comLr1/2}
\end{equation}
where $\beta_{ab}=-\beta_{ba}$ is defined by
$\beta_{ab} = \epsilon_{2ac}\epsilon{_1}{^c}_{b} -  \epsilon_{2bc}\epsilon{_1}{^c}_{a}$.
The commutators of two new NLSUSY transformations (\ref{newsusy})  on $\psi(x)$ and  ${e^{a}}_{\mu}(x)$ 
are GL(4R), i.e. new NLSUSY (\ref{newsusy}) is the square-root of GL(4R); 
\begin{equation}
[\delta_{\zeta_1}, \delta_{\zeta_2}] \psi
= \Xi^{\mu} \partial_{\mu} \psi,
\quad
[\delta_{\zeta_1}, \delta_{\zeta_2}] e{^a}_{\mu}
= \Xi^{\rho} \partial_{\rho} e{^a}_{\mu}
+ e{^a}_{\rho} \partial_{\mu} \Xi^{\rho},
\label{com1/2-e}
\end{equation}
where 
$\Xi^{\mu} = 2i (\bar{\zeta}_2 \gamma^{\mu} \zeta_1)
      - \xi_1^{\rho} \xi_2^{\sigma} e{_a}^{\mu}
      (\partial_{[\rho} e{^a}_{\sigma]})$.
They show the closure of the algebra. 
SGM action (\ref{SGM}) is invariant at least under\cite{st2}
\begin{equation}
[{\rm global\ NLSUSY}] \otimes [{\rm local\ GL(4,R)}] \otimes [{\rm local\ Lorentz}],  \\
\label{sgmsymm}
\end{equation}
which is isomorphic to SP whose single irreducible representation with N=10 gives 
the group theoretical description of SGM\cite{ks2}.  \par

The linearization of such a theory with a high nonlinearity is interesting  and 
inevitable to extract physics.  From the experience of the linearization of V-A model in flat spacetime 
we expect that we can carry out the linearization exactly and that we can obtain an equivalent 
local field theory which is renormalizable  and describes the observed low energy (SM) physics.   \\
The flat space linearization of N=1 V-A model has been carried out and proved that N=1 V-A model 
is equivalent to N=1 scalar supermultiplet\cite{ikruz} ${\it or}$ N=1 axial vector gauge supermultiplet 
of linear SUSY\cite{stt2}. \\
As a flat space exercise for the  extended SGM linerization, we have carried out the linearization of  
N=2 V-A model and  shown that it is equivalent to the spontaneously broken N=2 linear SUSY  
${\it vector \ J^{P}=1^{-}}$ gauge supermultiplet model with  SU(2) structure\cite{stt1}. 
Interestingly  SU(2) algebraic gauge structure of the electroweak standard model(SM) 
may be explained for the first time provided that the electroweak gauge bosons are  the composite  fields 
of this(SGM) type in the low energy.  \\
In these works the linearization are carried out by using the superfield formalism 
and/or by the heuristic and intuitive arguments 
on the relations between the component fields of LSUSY and NLSUSY. 
In either case it is crucial to discover  the SUSY invariant relations  which connect the supermultiplets 
of L and NL theories and reproduce the SUSY transformations.   \\
In abovementioned cases of the global SUSY in flat spacetime  the SUSY invariant relations are obtained 
straightforwardly, for L and NL supermultiplets are well undestood and the algebraic structures are the same SP.    \par
The situation  is rather different in SGM, for (i) the supermultiplet structure of 
the linearized theory of SGM is unknown except it is expected to be a broken SUSY SUGRA-like theory 
containing graviton and a (massive) spin 3/2 field as  dynamical d.o.f. and 
(ii) the algebraic structure (\ref{sgmsymm}) is changed into  SP.       \\
Therefore  by the heuristic arguments and referring to SUGRA  we discuss for the moment 
the linearization of N=1 SGM.          \par
At first, we assume  faithfully to SGM scenario that; \\
(i) the linearized theory should contain the  spontaneously broken ${\it global}$ (at least) SUSY  \\
(ii) graviton is an elementary field(not composite of superons coresponding to the vacuum of the Clifford algebra) 
in both L and NL theories   \\
(iii) the NLSUSY supermultiplet of SGM ($e{^a}_{\mu}(x)$, $\psi(x)$)  should be connected 
to the composite supermultiplet 
(${\tilde e}{^a}_{\mu}(e(x), \psi(x))$, ${\tilde \lambda}_{\mu}(e(x), \psi(x))$)  
for elementary graviton field and a composite (massive) spin 3/2 field of the SUGRA-like linearized theory. \par
From these assumptions and following the arguments performed in the flat space cases  we require that 
the SUGRA gauge transformation\cite{fvfdz} 
with the global spinor parameter ${\zeta}$ 
should hold for the supermultiplet (${\tilde e}{^a}_{\mu}(e, \psi)$, ${\tilde \lambda_{\mu}(e, \psi)}$)  
of the (SUGRA-like) linearized theory, i.e.,  \\
\begin{equation}
\delta {\tilde e}{^a}_{\mu}(e, \psi)  
      = i\kappa \bar{ \zeta} \gamma^{a} {\tilde \lambda_{\mu}(e, \psi)}, 
\label{sugral-2}
\end{equation} 
\begin{equation}
\delta {\tilde \lambda}_{\mu}(e, \psi)  
      = {2 \over \kappa}D_{\mu}{ \zeta}  
      = -{i \over \kappa}{{\tilde \omega(e, \psi)}_{\mu}}^{ab}\sigma_{ab}{ \zeta}, 
\label{sugral-3/2} 
\end{equation} 
where  $\sigma^{ab} = {i \over 4}[\gamma^a, \gamma^b]$, 
$D_{\mu}=\partial_{\mu}-{i \over 2} {{\omega}_{\mu}}^{ab}(e, \psi)\sigma_{ab}$, ${ \zeta}$ is a 
global spinor parameter and the variations in the left-hand side are induced by NLSUSY  (\ref{newsusy}). \par
We put the following  SUSY invariant relations which connect $e^{a}{_\mu}$ to ${\tilde e}{^a}_{\mu}(e, \psi)$;
\begin{equation}
{\tilde e}{^a}_{\mu}(e, \psi) = { e}{^a}_{\mu}(x).     
\label{relation-2}
\end{equation} 
This relation (\ref{relation-2}) is the assumption (ii) and holds simply the metric conditions.
Consequently the following covariant relation  is obtained by substituting  (\ref{relation-2}) 
into (\ref{sugral-2}) and  computing the variations under (\ref{newsusy})\cite{sts}; 
\begin{equation}
{\tilde \lambda}_{\mu}(e, \psi)  
      = \kappa \gamma_{a} \gamma^{\rho} \psi(x) \partial_{[\rho} e{^a}_{\mu]}.    
\label{relation-3/2} 
\end{equation} 
(As discussed later these should may be considered as the leading order of the expansions in $\kappa$ of 
SUSY invariant relations. The expansions terminate with $(\psi)^{4}$.)
Now we see LSUSY transformation 
%
%(\ref{sugral-2}) and  (\ref{sugral-3/2}) 
%
induced by (\ref{newsusy}) on the (composite) supermultiplet 
(${\tilde e}{^a}_{\mu}(e, \psi)$, ${\tilde \lambda}_{\mu}(e, \psi$)).    \\
The LSUSY transformation  on $\tilde e{^a}_{\mu}$ becomes  as follows. 
The left-hand side of (\ref{sugral-2}) gives
\begin{equation}
\delta {\tilde  e}{^a}_{\mu}(e, \psi) = \delta^{NL} {e^{a}}_{\mu}(x) 
= i \kappa^{2} (\bar{\zeta}{\gamma}^{\rho}\psi(x))\partial_{[\rho} {e^{a}}_{\mu]}(x). 
\label{susysgm-2} 
\end{equation} 
While substituting (\ref{relation-3/2}) into the righ-hand side of  (\ref{sugral-2}) we obtain 
\begin{equation}
i \kappa^{2} (\bar{\zeta}{\gamma}^{\rho}\psi(x))\partial_{[\rho} {e^{a}}_{\mu]}(x) + \cdots({\rm extra \ terms}).
\label{susysgm-3/2} 
\end{equation} 
These results show that  (\ref{relation-2}) and (\ref{relation-3/2}) are not  SUSY invariant relations 
and reproduce (\ref{sugral-2}) with unwanted extra terms which should be identified with the auxirialy 
fields. 
The commutator of the two LSUSY transformations induces GL(4R) with the field dependent parameters as follows;
\begin{equation}
[\delta_{\zeta_1}, \delta_{\zeta_2}]{\tilde  e}{^a}_{\mu}(e, \psi) 
= \Xi^{\rho} \partial_{\rho} {\tilde  e}{^a}_{\mu}(e, \psi)    
+ {\tilde  e}{^a}_{\rho}(e, \psi)\partial_{\mu} \Xi^{\rho},
\label{susysgmcom-2}
\end{equation}
where 
$\Xi^{\mu} = 2i (\bar{\zeta}_2 \gamma^{\mu} \zeta_1)
      - \xi_1^{\rho} \xi_2^{\sigma} e{_a}^{\mu}
      (\partial_{[\rho} e{^a}_{\sigma]})$.

On  ${\tilde \lambda}_{\mu}(e, \psi)$, the left-hand side of (\ref{sugral-3/2}) becomes 
apparently rather complicated; 
\ba
\delta {\tilde \lambda}_{\mu}(e, \psi)  
\A = \A {\kappa } \delta( \gamma_{a} \gamma^{\rho} \psi(x) \partial_{[\rho} e{^a}_{\mu]}) \nonu
\A = \A {\kappa } \gamma_{a}[ \delta^{NL}\gamma^{\rho} \psi(x) \partial_{[\rho} e{^a}_{\mu]} + 
  \gamma^{\rho} \delta^{NL} \psi(x) \partial_{[\rho} e{^a}_{\mu]} +
  \gamma^{\rho}  \psi(x) \partial_{[\rho} \delta^{NL} e{^a}_{\mu]}]. 
\label{susysgm-3/2} 
\ea
However the commutator of the two LSUSY transformations induces the similar GL(4,R);
\begin{equation}
[\delta_{\zeta_1}, \delta_{\zeta_2}]{\tilde \lambda}_{\mu}(e, \psi)  
= \Xi^{\rho} \partial_{\rho} {\tilde \lambda}_{\mu}(e, \psi)  
+ {\tilde \lambda}_{\rho}(e, \psi)\partial_{\mu} \Xi^{\rho}.  
\label{susysgmcom-3/2}
\end{equation}                  
These results indicate that it is necessary to generalize  (\ref{sugral-2}), (\ref{sugral-3/2}) and 
(\ref{relation-3/2})  for obtaining SUSY invariant relations and for the closure of the algebra.
%
%
% the algebra on the linearized field closes and  the initial new NL SUSY 
%structure of SGM is maintained  on the linearized supermultiplet (,i.e. disappointedly the  commutators 
%does not induce  super-Poincar\'e symmetry) 
%provided that the relations (\ref{relation-2}) and (\ref{relation-3/2}) 
%and SUGRA transformation (\ref{sugral-2}) are respected.  
%
%This shows that the left-hand side of  (\ref{sugral-2}), i.e. the right-hand side of (\ref{susysgm-3/2}) 
%is effectively a square-root of GL(4R).  
%
%
%
Furthermore  due to the complicated expression of LSUSY (\ref{susysgm-3/2}) which makes the physical and 
mathematical structures are obscure, we can hardly guess a linearized invariant action 
which is equivalent to SGM.       \par
Now we generalize the linearization by considering the auxirialy fields 
such that  LSUSY transformation on the linearized fields 
induces SP transformation.   \par
%
%
%
%As for the commutator of LSUSY transformation it generates GL(4R) 
%with the same field dependent parameters. \par
%
%
By comparing (\ref{sugral-3/2}) with (\ref{susysgm-3/2}) we understand that the local Lorentz transformation  
plays a crucial role. 
As for the local Lorentz transformation on the linearized asymptotic fields corresponding 
to the observed particles (in the low energy), 
it is natural to take (irrespective of (\ref{newlorentz})) the following forms   \\ 
\begin{equation}
\delta_L \tilde \lambda_{\mu}(x) = - {i \over 2} { \epsilon}_{ab}
      \sigma^{ab} \tilde \lambda_{\mu}(x),     \quad
\delta_L \tilde { e^{a}}_{\mu}(x) = {\epsilon}{^a}_b \tilde e{^b}_{\mu}, 
\label{lorentz}
\end{equation}
where $\epsilon_{ab} = (1/2)\epsilon_{[ab]}(x)$ is a local parameter.   
%
%
%
%The relation between (\ref{newlorentz}), i.e. the Lorentz invariance encoded 
%geometrically in  SGM space-time,  and (\ref{lorentz}), i.e.  the  Lorentz invariance 
%defined on the (composite) asymptotic field in Riemann space-time, is not known. 
%
%
%
In SGM  the local Lorentz transformations  (\ref{Lrw}) 
and (\ref{newlorentz}), 
i.e. the local Lorentz invariant gravitational interaction of superon, 
are introduced   by the geomtrical arguments in SGM spacetime\cite{st2} following  EGRT.  
While in SUGRA theory the local Lorentz  transfomation  invariance  (\ref{lorentz}) 
is  realized as usual by introducing  the Lorentz spin connection ${\omega^{ab}}{_\mu}$ . 
And the LSUSY transformation is defined successfully by the (Lorentz) covariant 
derivative containing the spin connection $\tilde \omega^{ab}{_\mu}(e,\psi)$ as seen in (\ref{sugral-3/2}), 
which causes the super-Poincar\'e algebra on the commutator of SUSY and is convenient for 
constructing the invariant action. 
Therefore in the linearized (SUGRA-like) theory the local Lorentz  transformation  invariance is expected 
to be realized as usual by defining (\ref{lorentz}) and introducing the Lorentz spin connection 
$\omega^{ab}{_\mu}$.
We investigate how the spin connection $\tilde\omega^{ab}{_\mu}(e,\psi)$ appears 
in the linearized (SUGRA-like) theory through  the linearization process. 
This is also crucial for constructing a nontrivial (interacting) linearized action 
which has manifest invariances.  \par  
We discuss the Lorentz covariance of the  transformation by comparing (\ref{susysgm-3/2}) with 
the right-hand side of (\ref{sugral-3/2}). 
The direct computation of (\ref{sugral-3/2}) by using the relations 
(\ref{relation-2}) 
and (\ref{relation-3/2}) under (\ref{newsusy}) produces complicated redundant terms 
as read off from (\ref{susysgm-3/2}). 
The local Lorentz invariance of the linearized theory may become ambiguous and lose the manifest invariance. \\
For a simple  restoration of the manifest local Lorentz invariance 
we survey the possibility that such redundant terms may be adjusted  by  the d.o.f of 
the auxiliary fields in the linearized supermultiplet.
As for the auxiliary fields it is necessary for the closure of the off-shell superalgebra 
to include the equal number of the fermionic and the bosonic d.o.f. in the linearized supermultiplet. 
As new NLSUSY is a global symmetry, ${\tilde \lambda}_{\mu}$ has 16 fermionic d.o.f.. 
Therefore at least 4 bosonic d.o.f. must be added to the off-shell SUGRA supermultiplet 
with 12 d.o.f.\cite{swfv} and a vector field  may be a simple candidate.  \par
However, counting the bosonic d.o.f. present in the redundant terms corresponding to 
$\tilde\omega^{ab}{_\mu}(e,\psi)$, 
we may need a bigger supermultiplet  e.g. $16 + 4 \cdot 16 = 80$  d.o.f., to carry out the linearization, 
in which case a rank-3 tensor $\phi_{\mu\nu\rho}$ 
and a rank-2 tensor-spinor $\lambda_{\mu\nu}$ may be candidates for the auxiliary fields.     \par
Now we consider  the  simple modification of SUGRA transformations(algebra) by adjusting 
the (composite) structure of the (auxiliary) fields.  
We take, in stead of (\ref{sugral-2}) and  (\ref{sugral-3/2}), 
\begin{equation}
\delta {\tilde e}{^a}_{\mu}(x) 
      = i\kappa \bar{\zeta} \gamma^{a} {{\tilde \lambda}_{\mu}(x)} + \bar{\zeta}{\tilde \Lambda}{^a}_{\mu}, 
\label{newsugral-2}
\end{equation} 
\begin{equation}
\delta {\tilde \lambda}_{\mu}(x) 
      = {2 \over \kappa}D_{\mu}\zeta + {\tilde \Phi}_{\mu}\zeta  
      = -{i \over \kappa}\tilde \omega^{ab}{_\mu}\sigma_{ab}\zeta + {\tilde \Phi}_{\mu}\zeta , 
\label{newsugral-3/2} 
\end{equation} 
where  $\tilde \Lambda{^a}_{\mu}$ and  $\tilde \Phi_{\mu}$ represent auxiliary fields 
which are functionals of $e^{a}{_\mu}$ and $\psi$. 
We need $\tilde \Lambda{^a}_{\mu}$ term in (\ref{newsugral-2}) to alter (\ref{susysgm-2}), 
(\ref{susysgmcom-2}), (\ref{susysgm-3/2}) and  (\ref{susysgmcom-3/2}) 
toward that of super-Poincar\'e algebra of SUGRA.  
We attempt the restoration of the manifest local Lorentz invariance order by order by adjusting 
$\tilde \Lambda{^a}_{\mu}$ and  $\tilde \Phi_{\mu}$. 
In fact, the Lorentz spin connection  ${\omega}^{ab}{_\mu}(e)$(i.e. the leading order terms of 
$\tilde\omega^{ab}{_\mu}(e, \psi)$) of (\ref{newsugral-3/2}) is reproduced by taking the following one  
\begin{equation}
\tilde \Lambda{^a}_{\mu} = {\kappa^{2} \over 4}[ ie_{b}{^\rho}\partial_{[\rho}e{^b}_{\mu]}\gamma{^a}\psi 
- \partial_{[\rho}e_{\mid b \mid \sigma]}e{^b}_{\mu}\gamma^{a}\sigma^{\rho\sigma}\psi ], 
\label{auxlambda-1}
\end{equation} 
where (\ref{susysgmcom-2}) holds.
Accordingly $\tilde \lambda_{\mu}(e,\psi)$ 
is determined up to the first order in $\psi$ as follows; 
\begin{equation}
\tilde \lambda_{\mu}(e,\psi) 
= { 1 \over 2i\kappa}( i\kappa^{2}\gamma_{a} \gamma^{\rho} \psi(x) \partial_{[\rho} e{^a}_{\mu]}
- \gamma_{a}\tilde \Lambda{^a}_{\mu} ) = -{i \kappa \over 2}\omega^{ab}{_\mu}(e)\sigma_{ab}\psi,  
\label{lambda-o1}
\end{equation} 
which indicates the minimal Lorentz covariant gravitational interaction of superon. 
Sustituting  (\ref{lambda-o1}) into (\ref{newsugral-3/2}) we obtain the following new LSUSY transformation 
of $\tilde \lambda_{\mu}$(after Fiertz transformations) 
\ba
\delta {\tilde \lambda}_{\mu}(e,\psi) 
\A = \A -{i \kappa \over 2} \{ \delta^{NL}\omega^{ab}{_\mu}(e)\sigma_{ab}\psi + 
{{\omega}_{\mu}}^{ab}(e)\sigma_{ab} \delta^{NL}\psi \}  \nonu
\A = \A -{i \over {2 \kappa}} \omega^{ab}{_\mu}(e)\sigma_{ab} \zeta + 
{i \kappa \over 2} \{ \tilde \epsilon^{ab}(e,\psi)\sigma_{ab}\cdot{}\omega^{cd}{_\mu}
(e)\sigma_{cd}\psi + \cdots \}.
\label{varlambda-o1} 
\ea 
Remarkably the local Lorentz transformations  of  ${\tilde \lambda}_{\mu}(e,\psi)$ (,i.e. the second term)
with the field dependent antisymmetric parameters $\tilde \epsilon^{ab}(e, \psi)$ is induced 
in addition to the intended ordinary global SUSY transformation. 
This shows that (\ref{lambda-o1}) is the SUSY invariant relations for $\tilde \lambda_{\mu}(e,\psi)$, 
for the SUSY transformation of (\ref{lambda-o1}) gives the right hand side of (\ref{newsugral-3/2}) 
with the extra terms. 
Interestingly the commutator of the two LSUSY transformations 
on (\ref{lambda-o1}) induces GL(4R); 
\begin{equation}
[\delta_{\zeta_1}, \delta_{\zeta_2}]{\tilde \lambda}_{\mu}(e, \psi)  
= \Xi^{\rho} \partial_{\rho} {\tilde \lambda}_{\mu}(e, \psi)  
+ \partial_{\mu} \Xi^{\rho}{\tilde \lambda}_{\rho}(e, \psi),  
\label{varlambda-o1-comm}
\end{equation}                   
where $\Xi^{\rho}$ is the same field dependent parameter as given in (\ref{susysgmcom-2}).
(\ref{susysgmcom-2}) and (\ref{varlambda-o1-comm}) show the closure of the algebra 
on SP algebra provided that the SUSY invariant relations (\ref{relation-2}) and (\ref{lambda-o1}) 
are adopted. 
These phenomena coincide with SGM scenario\cite{ks1}\cite{ks2} from the algebraic point of view, 
i.e. they are the superon-graviton composite (eigenstates) corresponding to the  linear representations 
of SP algebra. 
As for the redundant higher order terms in (\ref{varlambda-o1})  
we can  adjust them by considering the modified spin connection $\tilde\omega^{ab}{_\mu}(e, \psi)$ 
particularly with the contorsion terms and by recasting them 
in terms of (the auxiliary field d.o.f.) 
$\tilde \Phi_{\mu}(e,\psi)$. 
In fact,  we found that the following supermultiplet containing 
160 (= 80 bosonic + 80 fermionic) d.o.f. may be 
the supermultiplet of the SUGRA-like LSUSY theory which is equivalent to SGM;     \\
for 80 bosonic d.o.f.
\ba 
\A \A
[ \ \tilde e{^a}_{\mu}(e,\psi), a_{\mu}(e,\psi), b_{\mu}(e,\psi), M(e,\psi), N(e,\psi),  \nonumber   \\  
\A \A 
A_{\mu}(e,\psi), B_{\mu}(e,\psi), A{^a}_{\mu}(e,\psi), B{^a}_{\mu}(e,\psi), A{^{[ab]}}_{\mu}(e,\psi) \ ]
\label{80bosons}
\ea
and for 80 fermionic d.o.f.  \\
\ba 
[ \ \tilde{\lambda}{}_{\mu\alpha}(e,\psi), \  \tilde{\Lambda}{^a}_{\mu\alpha}(e,\psi) \ ],
\label{80fermions}
\ea
where $\alpha=1,2,3,4$ are indices for  Majorana spinor. The gauge d.o.f. of 
the local GL(4R) and the local Lorentz of the vierbein are subtracted. 
Note that the second line of (\ref{80bosons}) is equivalent to an auxiliary field with spin 3.    \\
The ${\it a \ priori}$ gauge invariance for  $\tilde \lambda_{\mu\alpha}(e,\psi)$ is not necessary for massive case\cite{f}
corresponding to the spontaneous SUSY breaking. 
For it is natural to suppose that the equivalent linear theory 
may be a coupled system of graviton and massive spin 3/2  with the spontaneous global SUSY breaking, 
which may be an analogue obtained by the super-Higgs mechanism 
in the spontaneous local SUSY breaking of N=1 SUGRA\cite{dz}.   \\
By continuing the heuristic arguments order by order referring to the familiar SUGRA supermultiplet 
we find the following SUSY invariant relations up to $O(\psi^{2})$: 
\ba 
\tilde e{^a}_{\mu}(e,\psi) 
\A=\A 
e{^a}_{\mu}, 
\label{compo-2} \\
%\ea
%
%\ba 
\tilde \lambda_{\mu}(e,\psi) 
\A=\A 
-i\kappa(\sigma_{ab}\psi)\omega^{ab}{}_{\mu},    
\label{compo-3/2} \\
%\ea
%
%
%\ba 
\tilde{\Lambda}{}^a{}_{\mu}(e,\psi)
\A=\A
\frac{\kappa^2}{2}\epsilon^{abcd}(\gamma_5\gamma_d\psi)\omega_{bc\mu}, 
\label{compo-5/2} \\
%\ea
%
%%
%\ba
A_{\mu}(e,\psi)
\A=\A 
\frac{i\kappa^2}{4}
[(\bar{\psi}\gamma^{\rho}\partial_{\rho}\tilde{\lambda}_{\mu})
-(\bar{\psi}\gamma^{\rho}\tilde{\lambda}_a)\partial_{\mu}e^a{}_{\rho}
-(\bar{\tilde{\lambda}}_{\rho}\gamma^{\rho}\partial_{\mu}\psi)]  \nonumber \\
\A \A
+\frac{\kappa^3}{4}
[(\bar{\psi}\sigma^{a\rho}\gamma^{b}\partial_{\rho}\psi)
(\omega_{\mu ba}+\omega_{ab\mu}) 
+(\bar{\psi}\sigma^{ab}\gamma^{c}\partial_{\mu}\psi)\omega_{cab} ] \nonumber \\
\A \A
+\frac{\kappa^2}{8}(\bar{\tilde{\lambda}}_{\mu}\sigma_{ab}\gamma^{\rho}\psi)\omega^{ab}{}_{\rho},  
\label{compo-Amu} \\
%\ea
%
%\ba
B_{\mu}(e,\psi)
\A=\A
\frac{i\kappa^2}{4}
[-(\bar{\psi}\gamma_5\gamma^{\rho}\partial_{\rho}\tilde{\lambda}_{\mu})
+(\bar{\psi}\gamma_5\gamma^{\rho}\tilde{\lambda}_a)\partial_{\mu}e^a{}_{\rho}
-(\bar{\tilde{\lambda}}_{\rho}\gamma_5\gamma^{\rho}\partial_{\mu}\psi)]   \nonumber \\
\A \A
+\frac{\kappa^3}{4}
[(\bar{\psi}\gamma_5\sigma^{a\rho}\gamma^{b}\partial_{\rho}\psi)
(\omega_{\mu ba}+\omega_{ab\mu}) 
+(\gamma_5\sigma^{ab}\gamma^{c}\partial_{\mu}\psi)\omega_{cab} ] \nonumber \\
\A \A
+\frac{\kappa^2}{8}(\bar{\tilde{\lambda}}_{\mu}\gamma_5\sigma_{ab}\gamma^{\rho}\psi)\omega^{ab}{}_{\rho}, 
\label{compo-Bmu} \\
%\ea
%
%%
%\ba
A^a{}_{\mu}(e,\psi)
\A=\A
\frac{i\kappa^2}{4}
[(\gamma^{\rho}\gamma^a\partial_{\rho}\tilde{\lambda}_{\mu})
-(\gamma^{\rho}\gamma^a\tilde{\lambda}_b)\partial_{\mu}\tilde{e}{}^b{}_{\rho}
+(\bar{\tilde{\lambda}}_{\rho}\gamma^a\gamma^{\rho}\partial_{\mu}\psi)]   \nonumber \\
\A \A
+\frac{\kappa^3}{4}
[-(\bar{\psi}\sigma^{b\rho}\gamma^a\gamma^{c}\partial_{\rho}\psi)
(\omega_{\mu cb}+\omega_{bc\mu})
-(\gamma^{bc}\sigma^a\gamma^{d}\partial_{\mu}\psi)\omega_{dbc} ]   \nonumber \\
\A \A
-\frac{\kappa^2}{8}(\bar{\tilde{\lambda}}_{\mu}\sigma_{bc}\gamma^a\gamma^{\rho}\psi)\omega^{ab}{}_{\rho}, 
\label{compo-Aamu} \\
%\ea
%
%
%\ba
B^a{}_{\mu}(e,\psi)
\A=\A 
\frac{i\kappa^2}{4}
[(\bar{\psi}\gamma_5\gamma^{\rho}\gamma^a\partial_{\rho}\tilde{\lambda}_{\mu})
-(\gamma_5\gamma^{\rho}\gamma^a\tilde{\lambda}_b)\partial_{\mu}\tilde{e}{}^b{}_{\rho}
+({\tilde{\lambda}}_{\rho}\gamma_5\gamma^a\gamma^{\rho}\partial_{\mu}\psi)]    \nonumber \\
\A \A
+\frac{\kappa^3}{8}
[-(\bar{\psi}\gamma_5\sigma^{b\rho}\gamma^a\gamma^{c}\partial_{\rho}\psi)
(\omega_{\mu cb}+\omega_{bc\mu})
-(\bar{\psi}\gamma_5\sigma^{bc}\gamma^a\gamma^{d}\partial_{\mu}\psi)\omega_{dbc} ]   \nonumber \\
\A \A
-\frac{\kappa^2}{8}(\bar{\tilde{\lambda}}_{\mu}\gamma_5\sigma_{bc}\gamma^a\gamma^{\rho}\psi)\omega^{ab}{}_{\rho},  
\label{compo-Bamu} \\
%\ea
%
%%
%\ba
A^{[ab]}{}_{\mu}(e,\psi)
\A=\A
\frac{i\kappa^2}{2}
[(\bar{\psi}\gamma^{\rho}\sigma^{ab}\partial_{\rho}\tilde{\lambda}_{\mu})
-(\bar{\psi}\gamma^{\rho}\sigma^{ab}\tilde{\lambda}_c)\partial_{\mu}\tilde{e}{}^c{}_{\rho}
+(\bar{\tilde{\lambda}}_{\rho}\sigma^{ab}\gamma^{\rho}\partial_{\mu}\psi)]       \nonumber \\
\A \A
-\frac{\kappa^3}{2}
[(\bar{\psi}\sigma^{c\rho}\sigma^{ab}\gamma^{d}\partial_{\rho}\psi)
(\omega_{\mu dc}+\omega_{cd\mu})
+(\bar{\psi}\sigma^{cd}\sigma^{ab}\gamma^{e}\partial_{\mu}\psi)\omega_{ecd} ]   \nonumber \\
\A \A
-\frac{\kappa^2}{4}(\bar{\tilde{\lambda}}_{\mu}\sigma_{cd}\sigma^{ab}\gamma^{\rho}\psi)\omega^{ab}{}_{\rho}. 
\label{compo-Aabmu}
\ea
%
%
%where the contorsion term of SUGRA-type is harmful and excluded.
%
%
In fact we can show that the following LSUSY transformations on (\ref{80bosons}) and (\ref{80fermions}) 
inuced by  NLSUSY (\ref{newsusy}) close among them(80+80 linearized multiplet). 
We show  the explicit expressions of some of the LSUSY transformations up to  $O(\psi)$. 
\ba
\delta \tilde{e}{}^a{}_{\mu} 
\A=\A
i\kappa \bar{\zeta}\gamma^a\tilde{\lambda}_{\mu}
-\epsilon^a{}_b\tilde{e}{}^b{}_{\mu}
+\bar{\zeta}\tilde{\Lambda}^a{}_{\mu} , 
\label{lsusy-compo-2} \\
\delta \tilde{\lambda}{}_{\mu} 
\A=\A 
-\frac{i}{\kappa}(\sigma_{ab}\zeta)\omega^{ab}{}_{\mu}
+\frac{i}{2}\epsilon^{ab}(\sigma_{ab}\tilde{\lambda}_{\mu})    \nonumber \\
\A \A
+A_{\mu}\zeta +B_{\mu}(\gamma_5\zeta)+A^a{}_{\mu}(\gamma_a\zeta)
+B^a{}_{\mu}(\gamma_5\gamma_a\zeta)+A^{ab}{}_{\mu}(\sigma_{ab}\zeta),
\label{lsusy-compo-3/2} \\
\delta \tilde{\Lambda}{^a}_{\mu}
\A=\A
\frac{1}{2}\epsilon^{abcd}(\gamma_5\gamma_d\zeta)\omega_{bc\mu},
\label{lsusy-compo-5/2} \\
%\ea
%
%\ba
\delta A{}_{\mu}
\A=\A
-\frac{1}{8}\left[
	i(\bar{\zeta}\gamma^{\rho}D_{\rho}\tilde{\lambda}_{a})\tilde{e}{}^a{}_{\mu}
	+3i(\bar{\zeta}\gamma^{a}D_{\mu}\tilde{\lambda}_{a})
	+2(\bar{\zeta}\sigma^{\nu\rho}\gamma_{\mu}D_{\nu}\tilde{\lambda}_{\rho})
	\right]
	\nonumber \\
\A \A
-\frac{1}{4\kappa}
	\left[
	3(\bar{\zeta}D_{\mu}\tilde{\Lambda}^a{}_{a})
	+i(\bar{\zeta}\sigma^{ab}D_{\mu }\tilde{\Lambda}_{ab})
	+i(\bar{\zeta}\sigma^{a\rho}D_{\rho }\tilde{\Lambda}_{(ab)})\tilde{e}{}^b{}_{\mu}
	\right]
	\nonumber \\
%\an \an
%-\frac{i}{4}(\bar{\zeta}\gamma^{\rho}\tilde{\lambda}_a)D_{\mu}\tilde{e}{}^a{}_{\rho} \nonumber \\
\A \A
+\frac{1}{16}
	\left[
	4i(\bar{\zeta}\gamma^{\rho}\tilde{\lambda}_a)\omega^a{}_{\rho\mu}
	+4(\bar{\zeta}\sigma^{bc}\gamma^a\tilde{\lambda}{}_{a})\omega_{bc\mu}
	-4(\bar{\zeta}\sigma^{a\rho}\gamma^b\tilde{\lambda}_{[\rho})\omega_{|ab|\mu ]}
	\right.
\nonumber \\
\A \A
%+\frac{1}{16}
	\hspace{1cm}\left.
	+4(\bar{\zeta}\sigma^{ab}\gamma^{c}\tilde{\lambda}_{a})\omega_{\mu cb}
	-3(\bar{\zeta}\sigma^{\rho}\gamma^{bc}\tilde{\lambda}_{[\rho})\omega_{|bc|\mu ]}
	+2i(\bar{\zeta}\sigma^{ab}\gamma_{\mu}\sigma^{cd}\tilde{\lambda}_{a})\omega_{cdb}
	\right]
	\nonumber \\
%\an \an
%+\frac{1}{8\kappa}
%		(\bar{\zeta}\gamma^{bc}\tilde{\Lambda}^a{}_{a})\omega_{bc\mu }
%	\nonumber \\
\A \A
-\frac{1}{8\kappa}
	\left[
	(\bar{\zeta}\gamma^{b}\gamma^a\sigma^{cd}{\tilde{\Lambda}}_{ab})
	+(\bar{\zeta}\sigma^{cd}\gamma^{b}\gamma^a{\tilde{\Lambda}}_{ab})
	\right]\omega_{cd\mu} , 
\label{lsusy-compo-1} \\
\delta A^a{}_{\mu}
\A=\A
\frac{1}{8}\left[
	-4i(\bar{\zeta}D_{\mu}\tilde{\lambda}^{a})
	+i(\bar{\zeta}\gamma^a\gamma^{\rho}D_{[\mu}\tilde{\lambda}_{\rho ]})
	+2(\bar{\zeta}\sigma^{\nu\rho}\gamma^a\gamma_{\mu}D_{\nu}\tilde{\lambda}_{\rho})
	\right]
	\nonumber \\
\A \A
+\frac{1}{4\kappa}
	\left[
	-i(\bar{\zeta}\sigma^{b\rho}\gamma^aD_{[\mu}\tilde{\Lambda}_{|b|\rho ]})
	-i(\bar{\zeta}\sigma^{\nu\rho}\gamma^aD_{\nu }\tilde{\Lambda}_{b\rho})\tilde{e}{}^b{}_{\mu}
	+(\bar{\zeta}\gamma^c\gamma^b\gamma^aD_{\mu}\tilde{\Lambda}_{bc})
	\right]
	\nonumber \\
\A \A
+\frac{1}{16}
	\left[
	-4i(\bar{\zeta}\gamma^{\rho}\gamma^a\tilde{\lambda}_b)\omega^b{}_{\rho\mu}
	-2(\bar{\zeta}\gamma^{\rho}\gamma^a\sigma^{bc}\tilde{\lambda}_{[\rho})\omega_{|bc|\mu ]}
	+2(\bar{\zeta}\gamma^a\sigma^{cd}\gamma^b\tilde{\lambda}_{b})\omega_{cd\mu}
	\right.
	\nonumber \\
\A \A
	\hspace{1cm}\left.
	+2(\bar{\zeta}\sigma^{cd}\gamma^a\gamma^b\tilde{\lambda}_{b})\omega_{cd\mu}
	+4(\bar{\zeta}\sigma^{b\rho}\gamma^a\gamma^c\tilde{\lambda}_{[\rho})\omega_{|bc|\mu ]}
	-4(\bar{\zeta}\sigma^{bc}\gamma^a\gamma^{d}\tilde{\lambda}_{b})\omega_{\mu dc}
	\right.
	\nonumber \\
\A \A
	\hspace{1cm}\left.
	-(\bar{\zeta}\gamma^a\gamma^{\rho}\sigma^{cd}\tilde{\lambda}_{[\rho})\omega_{|cd|\mu ]}
	-2(\bar{\zeta}\sigma^{bc}\gamma^a\gamma_{\mu}\sigma^{de}\tilde{\lambda}_{b})\omega_{dec}
	\right]
	\nonumber \\
\A \A
+\frac{1}{8\kappa}
	\left[
	(\bar{\zeta}\sigma^{b\rho}\gamma^a\sigma^{cd}\tilde{\Lambda}_{b[\rho})\omega_{|cd|\mu ]}
	-(\bar{\zeta}\sigma^{\nu\rho}\gamma^{a}\sigma^{bc}\tilde{\Lambda}_{\mu\nu})\omega_{bc\rho}
	+i(\bar{\zeta}\gamma^c\gamma^b\gamma^a\sigma^{de}\tilde{\Lambda}_{bc})\omega_{de\mu}
	\right.
	\nonumber \\
\A \A
	\hspace{1cm}\left.
	+i(\bar{\zeta}\sigma^{de}\gamma^c\gamma^b\gamma^a\tilde{\Lambda}_{bc})\omega_{de\mu}
	\right]
	\nonumber \\
\A \A
+\frac{\kappa}{2}(\bar{\zeta}D_{\mu}\Lambda^{\prime}{}^a) 
-\frac{\kappa}{4}(\bar{\zeta}\gamma^c\gamma^a\Lambda^{\prime}{}^b)\omega_{bc\mu}, 
\label{lsusy-compo-3} \\
\delta A^{[ab]}{}_{\mu}
\A=\A
\frac{1}{4}\left[
	-2i(\bar{\zeta}\gamma^{\rho}\sigma^{ab}D_{\rho}\tilde{\lambda}_{c})\tilde{e}{}^c{}_{\mu}
	+i(\bar{\zeta}\sigma^{ab}\gamma^{\rho}D_{\rho}\tilde{\lambda}_{c})\tilde{e}{}^c{}_{\mu}
	+i(\bar{\zeta}\sigma^{ab}\gamma^cD_{\mu}\tilde{\lambda}{}_{c}) 
	-2(\bar{\zeta}\sigma^{\nu\rho}\sigma^{ab}\gamma_{\mu}D_{\nu}\tilde{\lambda}_{\rho})
	\right]
	\nonumber \\
\A \A
+\frac{1}{2\kappa}
	\left[
	-(\bar{\zeta}\sigma^{ab}D_{\mu}\tilde{\Lambda}^{c}{}_{c}) 
	+i(\bar{\zeta}\sigma^{cd}\sigma^{ab}D_{\mu}\tilde{\Lambda}_{cd})
	+i(\bar{\zeta}\sigma^{c\rho}\sigma^{ab}D_{\rho }\tilde{\Lambda}_{(cd)})\tilde{e}{}^d{}_{\mu}
	\right]
	\nonumber \\
\A \A
+\frac{1}{8}
	\left[
	4i(\bar{\zeta}\gamma^{\rho}\sigma^{ab}\tilde{\lambda}_c)\omega^c{}_{\rho\mu}
	+4(\bar{\zeta}\sigma^{c\rho}\sigma^{ab}\gamma^d\tilde{\lambda}_{[\rho})\omega_{|cd|\mu ]}
	-4(\bar{\zeta}\sigma^{cd}\sigma^{ab}\gamma^{e}\tilde{\lambda}_{c})\omega_{\mu ed}
	\right.
\nonumber \\
\A \A
%+\frac{1}{32}
	\hspace{1cm}\left.
	-(\bar{\zeta}\sigma^{ab}\gamma^{\rho}\sigma^{de}\tilde{\lambda}_{[\rho})\omega_{|de|\mu ]}
	-2i(\bar{\zeta}\sigma^{cd}\sigma^{ab}\gamma_{\mu}\sigma^{ef}\tilde{\lambda}_{c})\omega_{efd}
	\right.
	\nonumber \\
\A \A
%+\frac{1}{32}
	\hspace{1cm}\left.
	-4i(\bar{\zeta}\sigma^{cd}\sigma^{ab}\sigma^{ef}\gamma_c\tilde{\lambda}{}_{d})\omega_{ef\mu}
	+2(\bar{\zeta}\sigma^{ef}\sigma^{ab}\gamma^d\tilde{\lambda}{}_{d})\omega_{ef\mu}
	\right]
	\nonumber \\
\A \A
+\frac{1}{4\kappa}
	\left[
	-4(\bar{\zeta}\sigma^{[b|c}\tilde{\Lambda}^{d|}{}_{d})\omega^{a]}{}_{c\mu}
	+i(\bar{\zeta}\sigma^{ab}\sigma^{cd}\tilde{\Lambda}^{e}{}_{e})\omega_{cd\mu}
	-(\bar{\zeta}\sigma^{cd}\sigma^{ab}\sigma^{ef}{\tilde{\Lambda}}_{cd})\omega_{ef\mu}
	\right.
	\nonumber \\
\A \A
%-\frac{i}{32\kappa}
	\hspace{1cm}\left.
	-(\bar{\zeta}\sigma^{c\rho}\sigma^{ab}\sigma^{de}\tilde{\Lambda}_{(c\mu )})\omega_{de\rho}
	-2(\bar{\zeta}\sigma^{ef}\sigma^{cd}\sigma^{ab}{\tilde{\Lambda}}_{cd})\omega_{ef\mu}
	\right], 
\label{lsusy-compo-5}
\ea
where $\epsilon^{ab}$ is the Lorents parameter and we put 
$\epsilon^{ab}=\xi^{\rho}\omega^{ab}{}_{\rho}$. 
$\delta B_{\mu}$ and $\delta B{^a}_{\mu}$ are similar to $\delta A_{\mu}$ and $\delta A{^a}_{\mu}$ 
respectively and omitted for simplicity. 
In the right-hand side of (\ref{lsusy-compo-3}) and $\delta B{^a}_{\mu}$, 
the last terms contain $\Lambda^{\prime}{}^a{}_{\mu}$ which is defined by 
$\Lambda^{\prime}{}^a{}_{\mu}=-\epsilon^{abcd}\gamma_5\psi\omega_{bcd}$ . 
Note that $\Lambda^{\prime}{}^a{}_{\mu}$ is not the functional of 
the supermultiplet (\ref{80fermions}), 
so we may have to treat $\Lambda^{\prime}{}^a{}_{\mu}$ as new auxiliary field. 
However, if we put 
$\epsilon^{ab}=\epsilon^{ab}(\tilde{\lambda}{}_{\mu}, \tilde{\Lambda}{}^a{}_{\mu})$, e.g. 
$\epsilon^{ab}=\bar\zeta\gamma^{[a}\tilde{\lambda}{}^{b]}$, 
$\Lambda^{\prime}{}^a{}_{\mu}$ does not appear in 
the right-hand side of (\ref{lsusy-compo-3}) and $\delta B^{a}{_\mu}$. 
As a result, the LSUSY transformation 
on the supermultiplet (\ref{80bosons}) and (\ref{80fermions}) 
are written by using the supermultiplet itself 
at least at the leading order of superon $\psi$. 
The higher order terms remain to be studied.    
However we believe that we can obtain the complete linearized off-shell supermultiplets of 
the SP algebra by repeating the similar procedures (on the auxiliary fields) order by order 
which terminates with $\psi^{4}$.  It may be favorable that  10 bosonic auxiliary fields, for example 
${a_{\mu}(e,\psi), b_{\mu}(e,\psi), M(e,\psi), N(e,\psi)}$ are arbitrary  up now and available 
for the closure of the off-shell SP algebra in higher order terms.     \par
Finally we mention  the systematic linearization by using the superfield formalism 
applied to study the coupled system of V-A action with SUGRA\cite{lr}. 
We can define on such a coupled system 
a local spinor gauge symmetry which induces a super-Higgs mechanism\cite{dz} converting V-A field to 
the longitudinal component of massive spin 3/2 field. The consequent Lagrangian  
may be an analogue  that we have  anticipated in the composite picture but with the elementary spin 3/2 field. 
Developing the superfield formalism\cite{wb} 
on SGM spacetime may be crucial  for carrying out 
the linearization along the SGM composite scenario, especially for $N>1$.    \\
The linearization of SGM action (\ref{SGM}) with the extra dimensions, 
which gives another unification framework describing the observed particles as elementary fields, is open. 
And the linearization of SGM action for spin 3/2 N-G fermion field\cite{st3} (with extra dimensions) 
may be in the same scope.  \par
Now we summarize the results as follows: 
(i) Referring to SUGRA transformations we have obtained explicitly the SUSY invariant 
relations up to $O(\psi)^{2}$ and the corresponding new LSUSY transformations 
among 80+80 off-shell supermultiplet of LSUSY. 
(ii) The new LSUSY transformations on 80+80 linearized supermultiplet are different apparently 
from SUGRA transformations but close on super-Poincar\'e. 
(iii)It is interesting that the simple relation $\lambda_{\mu}=e^{a}{_\mu}\gamma_{a}\psi + \cdots $, 
which is sugested by the flat spacetime linearization, seems disfavour with the SGM linearization 
in our present method, so far. 
From the physical viewpoint what LSUSY SP may be to SGM in quantum field theory, 
what O(4) symmetry is to the relativistic hydrogen model in quantum mechanics. 
The complete linearization to all orders up to $O(\psi)^{4}$, 
which can be anticipated by the systematics emerging in the present study, needs specifications of 
the auxiliary fields and remains to be studied. The details will appear separately\cite{sts2}.

\vskip 15mm

The authors  would like to thank U. Lindstr\"om for the interest in our works and for bringing the reference 
to our attentionss. 
The work of M. Sawaguchi is supported in part by the research project of  High-Tech Research Center 
of Saitama Institute of Technology.

\newpage

%%%%%%%  References  %%%%%%%%%%%%%%%%%%%%%%%%%%%%%%%%%%%%%%%
%
\newcommand{\NP}[1]{{\it Nucl.\ Phys.\ }{\bf #1}}
\newcommand{\PL}[1]{{\it Phys.\ Lett.\ }{\bf #1}}
\newcommand{\CMP}[1]{{\it Commun.\ Math.\ Phys.\ }{\bf #1}}
\newcommand{\MPL}[1]{{\it Mod.\ Phys.\ Lett.\ }{\bf #1}}
\newcommand{\IJMP}[1]{{\it Int.\ J. Mod.\ Phys.\ }{\bf #1}}
\newcommand{\PR}[1]{{\it Phys.\ Rev.\ }{\bf #1}}
\newcommand{\PRL}[1]{{\it Phys.\ Rev.\ Lett.\ }{\bf #1}}
\newcommand{\PTP}[1]{{\it Prog.\ Theor.\ Phys.\ }{\bf #1}}
\newcommand{\PTPS}[1]{{\it Prog.\ Theor.\ Phys.\ Suppl.\ }{\bf #1}}
\newcommand{\AP}[1]{{\it Ann.\ Phys.\ }{\bf #1}}

\end{document}